\documentclass[a4paper]{article}
\usepackage[pdftex]{graphicx}
\pdfoutput=1
\usepackage{geometry}
\newcommand{\be}{\begin{equation}}
\newcommand{\ee}{\end{equation}}
\newcommand{\bd}{\begin{displaymath}}
\newcommand{\ed}{\end{displaymath}}
\newcommand{\bea}{\begin{eqnarray}}
\newcommand{\eea}{\end{eqnarray}}
\newcommand{\bi}{\begin{description}}
\newcommand{\ei}{\end{description}}
\newcommand{\bq}{\begin{quote}}
\newcommand{\eq}{\end{quote}}

\begin{document}
\bibliographystyle{unsrt}
\author{Alexander~Unzicker\\
        Pestalozzi-Gymnasium  M\"unchen\\[0.6ex]\\
{\small{\bf e-mail:}  alexander.unzicker@lrz.uni-muenchen.de}}

\author{
Alexander~Unzicker\\
         {\small Pestalozzi-Gymnasium M\"unchen, Germany}\\
           {\small aunzicker@web.de}}
\title{How to Determine the Probability of the Higgs Boson Detection}
\maketitle

\begin{abstract}

The Higgs boson is the most important, though yet undiscovered
ingredient of the standard model of particle physics.
Its detection is therefore one of the most important goals of high
energy physics that can guide future research in
theoretical physics. Enormous efforts have been undertaken
to prove the existence of the Higgs boson, and the physics
community is excitedly awaiting the restart of the Large Hadron Collider
at CERN.  But how sure can we be that the Higgs exits at all? 
The German philosopher Immanuel Kant recommended betting at such
controversial questions, and Stephen Hawking announced a \$100 bet
against the Higgs.
But seriously, online prediction markets, which are a generalized form
of betting, do provide the best possible probability estimates for 
future events. It is proposed that the scientific community uses this
platforms for evaluation. See also an online description 
{\em www.Bet-On-The-Higgs.com\/}.

\end{abstract}

\paragraph{A simple analogy and fundamental problems.}

The Higgs boson, predicted in 1964 by Peter Higgs \cite{Hig:64},
isnowadays considered the cornerstone of the
Standard Model of particle physics. Nobel award winner Leon Lederman called it the
`God particle', and as one of a few science problems, it has been publicizised
with the following analogy:

\begin{quote}
The British Prime Minister Margret Thatcher holds a cocktail party.
As she moves through the room, the cluster people around her, corresponding to the 
Higgs field, creates resistance to her movement, which appears as inertia or mass.
As single Higgs particle instead corresponds to a rumor 
which causes clusters of people standing close to each other.
\end{quote}

Though this story has also been told with John Major, 
Tony Blair and Gordon Brown, the detection of the Higgs particle is still 
missing. Since the old ideas of Mach \cite{Mach} and Dirac \cite{Dir:37},
 the question to the origin
of mass puzzles fundamental physics. In the context of the 
standard model, progress is expected with the detection of a scalar
boson like the Higgs which provides a mechanism for inertia 
mass. However, even if the Higgs should be detected, no prediction 
of mass ratios like $1836.15...$ (proton-electron) is expected.
Instead of 
not calculable masses, physics would be left with only
measurable couplings with the Higgs field. Nonetheless, there are few
other ideas in theoretical physics concerning the mass problem.

\paragraph{Experimental efforts.}

The energy at which the Higgs is expected ranges from 115 to 190
GeV \cite{Yao:06}. However, recent experiments at Tevatron have excluded the range
160-170 GeV on a $3 \sigma$-level \cite{fer:09}. While a heavy Higgs should quite
easily be detected at the Large Hadron Collider (LHC) at CERN, a mass
below 150 GeV would increase the background signal problem.
However, once it becomes fully operational, the best chances for the Higgs detection
rest on the Large Hadron Collider: after all, the Higgs detction was
the main goal for its construction.
LHC documentation videos claim that physics would be shocked by an
"earthquake" if the Higgs is not detected. With the restart of the LHC in 2009,
 the question becomes topical \cite{LHC:08}.
 
\paragraph{Public interest.} The amount of research funds spent for the
Higgs detection is extraordinary. A Superconducting Super Collider
project was cancelled in 1993 in the US, but also the three billion dollar LHC 
construction has raised criticism on high energy physics receiving
 unproportional funding. Thus besides the scientific interest of 
the high energy community, other physicists, scientists, taxpayers and
politicians may have a legitimate interest in the main purpose 
of the LHC. A central parameter to evaluate these efforts
is the probability of the Higgs detection. 

\paragraph{Existence or detection~?} 
A couple of unfortunate incidents, most importantly the short circuit 
damage on Sept 10th, 2008 caused a delay of the experiments
which hopefully restart in late 2009. Obviously, one has to distinguish
the probability of the existence of the Higgs from the probablility of 
its detection, which can be defined for a certain period only.
However, given that the LHC has all the capacities to cover the energy range
not yet excluded, sooner or later the Higgs must be detected
if it exists at all. It is a delicate task to interprete partial negative results.
Given that Tevatron results exclude a Higgs in the range from 160 to 170 GeV,
does that mean that the absolute probability for a lighter Higgs has increased
(as it was stated) or did the overall probability for the existence decrease~?
To carry to the extremes, can the non-discovery be a hint or even `proof'
the existence of a Higgs multiplett at higher energies~? Since 
supersymmetric theories usually do
not predict mass ranges, they seem to be `earthquake-safe' against falsification.
All these questions require reliable estimates of a priori probabilities
for the existence of a given particle.

\paragraph{The standard procedure: polling.} 

\bq
`Prediction is very difficult, especially about the future.' {\em Niels Bohr\/}
\eq

Elections are a well-known example where the probability is of public interest.
The outcome is usuallly forecasted by polls. This 
can be done with some success since elections follow a well-defined procedure and
due to the one man-one vote principle, do not need a weighting of the asked persons.
But how to do a poll on the Higgs particle detection~? Among the average population,
any poll would be completely useless, but even when asking
scientists or physicists, such a poll wouldn't generate a reliable result.
One has to ask the experts. But which experts~? CERN technicians, theoretical 
physicists which worry about the nature of mass, or Nobel award winners~?
An appropriate weighting of opinions is obviously impossible.
Then, elections are anonymous, and voters are not forced to comply with expectations
of the questioner. Scientists sometimes are, for an expert it could be 
embarrassing to state in public that the goal of a recent grant 
applicationan is unkikely.

\paragraph{Prediction markets.} 
An interesting alternative to polls are prediction markets \cite{pre}. They
work simlar to stock markets, a generalized form of betting.
 Even in the case of elections, they have been remarkably sucessful \cite{Eri:08}.
As stocks, prediction markets are anonymous. Most importantly, an opinion is
naturally weighted by the the amount of money one is willing to stake.
On the other hand, prediction markets do not consider themselves as 
gambling platforms, since the outcome is not determined 
primarily by chance but by the ability of the user to forsee 
developments. However, there is an obvios adjacency to the
suspicious sports betting business. In most countries, bets 
and gambling contracts do not constitute a legal committment,
thus it is not easy to place a real money bet on the discovery of the Higgs
boson. Though Stephen Hawking announced his \$100 bet against the Higgs,
no one can go to a court in case he's wrong (though his motives
are surely honest). \footnote{Another interesting science bet takes place
between John Horgan and Michio Kaku (http://www.longbets.org/12):
 The former is convinced that there will be
no Nobel award for a unified theory until 2020. The 
betting platform however requires a donation of a possible win.}
 However, the prediction market Intrade.com 
is located in Ireland where legislation does not inhibit operating, and
it is one of the few platforms  with scientific topics. Therefore, 
a brief explanation how to trade there is given in the next section.

\paragraph{How to trade the Higgs at a prediction market.}\footnote{A
online description is given at the site {\em www.Bet-On-The-Higgs.com\/}. 
There, every month the most optimistic bets on the Higgs are accepted for
US\$ 100.}

A prediction market like Intrade is based on the trade of contracts
on given events. The idea is that the actual price
is a measure of the probability that the event will happen.
If you belive for instance that the Higgs boson will be
discovered until the end of 2011, you may buy the
contract HIGGS.BOSON.DEC11. It is noteworthy that you may 
even {\em sell} contracts that you do not hold, in case you don't
believe the event will happen.

After opening an account and uploading a deposit with a credit card
one may start betting immediately. 
Contracts have a nominal value of \$10 corresponding to 100 points.
You may buy one of the contracts HIGGS.BOSON.DEC2010
or even HIGGS.BOSON.DEC2013 if you believe the discovery 
 will occur within the period \footnote{The deadline is the 
date of publication.} or sell, if you don't believe.
As in a stock market, there is a bid and an ask 
price.\footnote{{\em http://en.wikipedia.org/wiki/Ask\_price\/}}
If the event happens, the contract achieves the value 100, if not, zero.
This relates the current rate directly to the percent probability that the event will happen.

\paragraph{A concrete example.} 

After logging in, the keyword search `Higgs' among others shows the contract
HIGGS.BOSON.DEC11. Suppose that the Bid is 40 and the Ask is 44. If you believe in the
discovery, you may buy the contract immediately for 44 or place a buy 
order at 43. If you bought the contract for 44 and the particle
is discovered, the price rises to 100, your win is 56 points or 5.60 Dollars.
If you do not believe in the discovery by the end of 2011,
you may sell it right now for 40 and get 4.00 Dollars or place a sell order
for 42. If somebody accepts this 42 offer, you get 4.20 Dollars,
 but if the particle is discovered, you have to pay 10 Dollars to the
purchaser.
Therefore, you have to deposit this amount until the contract ends. There are moderate
fees for trading, usually fractions of a percent.

Figs.~1 and~2 show typical screenshots of an order book (offers not yet accepted)
and an position book (sold/bought contracts).

\begin{figure}
\includegraphics[width=14.0cm]{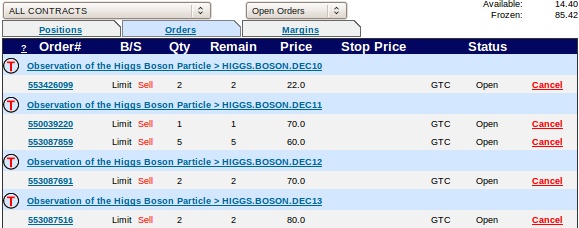}
\caption{Typical orders overview at intrade.com. }
\end{figure}
\begin{figure}
\includegraphics[width=14.0cm]{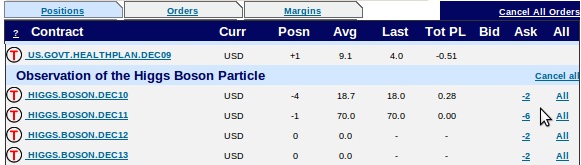}
\caption{Typical positions overview at intrade.com. }
\end{figure}

\paragraph{Summary and outlook.} 
Trading on prediction markets is an interesting procedure that
yields an approximate probability for the detection of the Higgs
boson. The value of the long-term contracts seems to be the 
best possible guess about the existence of the Higgs.
Of course, an appropriate volume (number of traded
contracts) is desirable to increase reliability, and at the
same time the trading volume could become a measure of importance 
the scientific community is willing to give to an open problem.
It would be interesting to extend the proposed method to other
expected discoveries like gravitational waves or extraterrestrial
life. On the other hand prediction markets can throw light on
speculative theories: who would seriously bet
on the discovery of micro black holes or supersymmetric particles~?
String theory, which has been acused of being non falsifiable any 
more \cite{Smo:06, Woi:06} would indeed have difficulties to define a contract about
the evidence to be discovered in its favor. 

Prediction markets could become an useful instrument for
research funding procedures. The German philosopher Immanuel Kant
said, the bet naturallydistinguishes between real
conviction and simple opinion. Prediction markets are a modern version
of such bets.

\end{document}